\documentclass[anonymous]{llncs}
\usepackage[T1]{fontenc}
\usepackage{graphicx}
\usepackage{amsmath}
\usepackage{amssymb}

\begin{document}
\title{A "Zero-Knowledge" Revocable Credential Verification Protocol Using Attribute-Based Encryption}
\titlerunning{Revocable Functional Credentials from ABE}
%
%
\author{Giovanni Bartolomeo\orcidID{0000-0002-7485-9884}}
\institute{CNIT, Italy}
\email{giovanni.bartolomeo@uniroma2.it}

%
\title{Revocable Anonymous Credentials from Attribute-Based Encryption}


%
%
\maketitle              
\begin{abstract}
 We introduce a credential verification protocol leveraging on Ciphertext-Policy Attribute-Based Encryption. The protocol supports anonymous proof of predicates and revocation through accumulators.

\keywords{Anonymous Credentials \and Attribute-Based Encryption \and Cryptography.}
\end{abstract}
\section{Introduction}
Anonymous credentials\footnote{For historical reasons, the title of the present paper hints at "Zero-Knowledge". We prefer the more accurate term "Anonymous credentials", although the protocol presented throughout the paper might be proved to actually be a Zero-Knowledge protocol.} have experimented a renewed interest during the very last few years due to the going to mainstream of various user "wallet" models. For example, an ongoing effort at IEFT is intended to promote a very recent efficient construction of the BBS signature \cite{cryptoeprint:2023/275} as a standard cryptographic primitive for the problem space of privacy preserving identity credentials. Historically, anonymous credentials were mostly built on special signature schemas. The prover, after obtaining a signature over a set of attributes from an issuer, is able to randomize it and proves in zero-knowledge its possession to a verifier, optionally revealing a subset of those attributes (so called \emph{selective disclosure}). The verifier is unable to determine which signature was used to generate the proof, removing any source of correlation (\emph{unlinkability}).

However, in practical contexts, selective disclosure is not the only desired feature. For example, a service may require that their users are over 18 years old and that they are based in one of the European Countries. In such a case, an \emph{anonymous proof of predicates} proving user's attribute \texttt{age} and \texttt{country} satisfying the following policy: 
\[
\texttt{age GT 18 AND country ONEOF \{Austria, Belgium,\dots, Sweden\}} 
\]
would be needed. Recent advances \cite{FCREDENTIALS} suggests it is relatively easy to build credential management systems supporting efficient anonymous proof of predicates by functional encryption: be given a ciphertext encoded using a policy, a prover can simply decrypt such a ciphertext to convince a verifier that she knows a key for a set of attributes that matches the policy. According to the authors of \cite{FCREDENTIALS}, such \emph{functional credentials} would subsume all known credentials, such as anonymous, delegatable, or attribute-based credentials.

In this paper we provide the following contributions:
\begin{enumerate}
    \item We directly build a functional credentials schema from the Ciphertext Policy Attribute-Based Encryption (CP-ABE) construct proposed in \cite{cryptoeprint:2008/290}, without incurring in few extra commitment steps introduced by the author of \cite{FCREDENTIALS} (originally required to leave their framework agnostic from any specific functional encryption schema). We consider this result important in practice for two reasons: first because many existing authentication protocols (such as HTTP and OAuth) relies on a three steps procedure (request-challenge-response), so avoiding any extra step would perfectly fit them.
    \item Revocation for functional credentials is still unclear and not investigated, while it is very relevant in real world applications. We propose to augment the above schema with an efficient anonymous revocation feature leveraging on the simple dynamic accumulator originally proposed in \cite{cryptoeprint:2008/539}.
    \item We achieve performance comparable to state of the art solutions without incurring in complex zero-knowledge proof algoriths, but solely relying on a consolidated attribute-based encryption schema. Again, this result is important in practice: for example, to quickly build credential systems with anonymous proof of predicates, developers can directly rely on the well-known OpenABE framework \cite{OABE} leveraging on the wide policy expressiveness this framework supports.
\end{enumerate}

\section{Preliminaries}
\subsection{Revocable Functional Credentials} \label{subsec:Revocable Functional Credentials}
The schema is adapted from \cite{FCREDENTIALS}.
A revocable functional credentials scheme for an attribute universe $\mathtt{\Omega}$ and a family of policies $\mathtt{\Phi}$ consists of the following probabilistic algorithms:
\begin{enumerate}
\item $\mathtt{MSK, MPK \leftarrow CKGen(1^\lambda)}$: The key generation algorithm takes input the security parameter $\mathtt{\lambda \in N}$ and outputs a key pair $\mathtt{(MSK, MPK)}$ of an issuer (master key pair).
\item $\mathtt{cred_i, MPK^\prime \leftarrow GrantCred(MSK, S)}$: The grant credential algorithm takes input the master secret key $\mathtt{MSK}$ and a non-empty set of attributes $\mathtt{S \subset \Omega}$ and outputs a credential $\mathtt{cred_i}$ with $\mathtt{i \in N}$ for the corresponding set of attributes. It also outputs an updated master public key $\mathtt{MPK^\prime}$
\item $\mathtt{b \leftarrow <ShowCred(MPK, cred_i, f),VrfyCred(MPK, f)>}$: $\mathtt{ShowCred}$ takes input the master public key $\mathtt{MPK}$, a credential $\mathtt{cred_i}$, and a policy $\mathtt{f} \in \Phi$; $\mathtt{VrfyCred}$ inputs the master public key $\mathtt{MPK}$ and a policy $\mathtt{f}$. At the end, $\mathtt{VrfyCred}$ outputs either $\mathtt{0}$ or $\mathtt{1}$.
\item $\mathtt{MPK^\prime \leftarrow Revoke(MPK,MSK,i)}$: takes input the master public key $\mathtt{MPK}$, the master secret key $\mathtt{MSK}$, the index $\mathtt{i}$ of the credential to be revoked and outputs an update master public key $\mathtt{MPK^\prime}$.

\end{enumerate}

By definition, for all $\mathtt{\lambda \in N}$, for all $\mathtt{(MSK, MPK) \in CKGen(1^\lambda)}$ for all $\mathtt{S \subset \Omega}$, for all $\mathtt{cred \in GrantCred(osk, S)}$, for all $\mathtt{f \in \Phi}$ such that $\mathtt{f(S) = 1}$, a functional credentials scheme 
\begin{itemize}
    \item is said \emph{correct} if it holds that 
\[\mathtt{Pr[1 \leftarrow <ShowCred(MPK, cred, f), VrfyCred(MPK, f)> ]=1}\]
    \item is said \emph{unforgeable} if, chosen an arbitrary policy $\mathtt{f}$, any adversary having access to all system issued credentials $\mathtt{cred_i}$ but the ones satisfying the policy (i.e, for each $\mathtt{cred_i}$ the adversary obtains $\mathtt{f(cred_i) \neq 1}$ ) has a negligible probability to succeed in the credential verification process.
    \item is said \emph{anonymous} if, arbitrarily chosen a policy $\mathtt{f}$ and two provers $\mathtt{P_0}$ and $\mathtt{P_1}$ owing credentials $\mathtt{cred_i}$ and $\mathtt{cred_j}$, both satisfying or not the policy (i.e., $\mathtt{f(cred_i)=f(cred_j)}$), any adversary acting as a verifier cannot distinguish between them.
\end{itemize}
Note that we leave off one optional feature described in the original paper (policy hiding).
\subsection{CP-WATERS-KEM}
The original schema reported in Section 5 of \cite{cryptoeprint:2008/290} (henceforth CP-WATERS-KEM) makes use of a bilinear group, defined as follows:

Let $G$ and $G_T$ be two multiplicative cyclic groups of prime order $p$. Let $g$ be a generator of $G$ and $e$ be a bilinear map: $e\colon G \times G \to G_T$. The bilinear map $e$ has the following properties:

\begin{enumerate}
    \item Bilinearity: for all $u, v \in G$ and $a, b \in Z_p$, we have $e(u^a,v^b) = e(u^b,v^a) = e(u,v)^{ab}$.
    \item Non-degeneracy: $e(g,g)\neq1$.
\end{enumerate}
If the group operation in $G$ and the bilinear map $e$ are both efficiently computable, $G$ is said a bilinear group.

Here, we consider CP-WATERS-KEM in its small universe construction, however, an extension to the large universe construction (reported in Appendix A of \cite{cryptoeprint:2008/290}) is straightforward. The schema consists of four algorithms:
\begin{enumerate}
    \item $\mathtt{(MPK, MSK) \leftarrow Setup(1^\lambda)}$: Using the security parameter $\lambda$, the algorithm outputs the master secret key $MSK$ and the master public key $MPK$, and publishes the $MPK$. 
    \begin{itemize}
        \item The algorithm chooses a group $G$ of prime order $p$ and generator $g$, random group elements $h_1, h_2, \dots, h_u$ (where $u$ is the maximum number of system attributes) and a bilinear pairing $e$ such that $e\colon G \times G \to G_T$. In addition, it chooses random exponents $\alpha, a \in Z_p$. 
        \item The public key is $g, g^a , e(g,g)^\alpha, h_1, h_2, \dots, h_u$ and the master secret key is $g^\alpha, \alpha, a$.
    \end{itemize}
    \item $\mathtt{SK=(K,L,\forall_{x \in S} \; K_x) \leftarrow KeyGen(MPK,MSK, S)}$: Key generation happens by taking as input the master secret key MSK and a set of attributes $S$ that describe the key. The output is a randomized secret decryption key.
    \begin{itemize}
	\item Chosen a random $t \in Z_p$ the algorithm simply generates and releases: $K= g^\alpha g^{at}$, $L = g^t$ and, for each $x \in S$, $K_x = h_x^t$  as the secret decryption key.
    \end{itemize}
    \item $\mathtt{(C=(C^\prime,\forall_{k \in [1,\dots\,l]} \; C_k), \mu)\leftarrow Encrypt(MPK, M^{l \times m}, \rho)}$: The algorithm takes as input an access structure $(M, \rho)$ and the public key $MPK$. $M$ is an $l \times m$ matrix, while $\rho$ is an injective function associating each row of $M$ to an attribute $\rho_k$ (i.e., $\rho_k=\rho(k) \in S$); note that in this construct one attribute is associated with at most one row (\cite{cryptoeprint:2008/290} proposes off-the-shelf techniques to cope with this limitation). The output is a random secret and the ciphertext.
    \begin{itemize}
	   \item Chosen a random \footnote{In order to use pseudo-randomness, the algorithm can take as input an optional input seed $u \in \{0, 1\}^k$ to a pseudo-random generator $PRG$. Later in the paper, we will use this feature to transform the ABKEM schema into a hybrid CCA-IND2 encryption schema.} vector $\vec{v}=(s,y_2,\dots,y_m)$ in $Z_p^m$ and being $M_k$ the k-th row of $M$, the algorithm computes $\lambda_k=v\cdot M_k$.
	\item Together with the access structure (M; $\rho$), the algorithm makes public the ciphertext: $C^\prime= g^s$ and $C_k=g^{a \lambda _k}h_{\rho_k}^{-s}$
	\item The algorithm computes the random secret $\mu=e(g,g)^{\alpha s}$ and keeps it private.
    \end{itemize}
    \item $\mathtt{\mu \leftarrow Decrypt(SK, C)}$: Dually, the decryption takes as input the ciphertext $C$ generated at step 3 and the secret key $SK$ generated at step 2. The output is the common secret $m=e(g,g)^{\alpha s}$ if and only if the set of attributes $S$ satisfies the access structure, or $null$ otherwise.
    \begin{itemize}
        \item 	For each $k$ such that $\rho_k \in S$ (i.e., consider only attributes in $S$), compute $\omega_k$ such that $\sum_k{\omega_k\lambda_k=s}$ (there could different sets of $\{ \omega_k \}$ satisfying this equation)
	\item Compute the random secret: $\mu=\frac{e(C^\prime,K)}{\prod_k{[e(C_k,L)e(C^\prime,K_{\rho_k})]^{\omega_k}}} = e(g,g)^{\alpha s}$
    \end{itemize}
\end{enumerate}

\section{Revocation}
Consider the CP-WATERS-KEM scheme whose implementatio was reported in the previous section. We implement a revocation scheme preventing decryption of ciphertext created after the key has been revoked. Worthy to be noted, this is not a general revocation scheme for ABE. Rather, it is a “forward revocation”, where only ciphertext generated after the actual revocation happens becomes hard to decrypt if the key has been revoked. To implement this kind of revocation, the original CP-WATERS-KEM scheme is slightly altered and combined with the cryptographic accumulator based on bilinear mappings described by Camenisch in \cite{cryptoeprint:2008/539}. 

The accumulator makes use of a set of indexes $\{i\}$ kept by the Authority and assigned to each released secret key. In the setup phase, the Authority initially creates an accumulator $acc_0=1$ and two public initially empty sets: $V=\{\}$ and $U=\{\}$, where $U$ is the set of all indexes $i$ that will be ever added to the accumulator (but may have been subsequently removed). The sequence $g^\gamma,\dots,g^{\gamma^n} ,g^{\gamma^{n+2}},\dots,g^{\gamma^{2n}}$ (but not $g^{\gamma^{n+1}}$) is made public by the Authority (e.g., as part of $MPK$). Appendix D of \cite{cryptoeprint:2008/539} suggests a possible technique to reduce the size of this sequence.
The mathematical definition of the accumulator is the following:
\[
acc_V=\prod_{j \in V}{g^{\gamma^{n+1-j}}}=g^{\frac{\sum_{j \in V}{\gamma^{n+1+i-j}}}{\gamma^i}}
\]
\[									
wit_i=\prod_{j \neq i}{g^{\gamma^{n+1+i-j}}}=g^{\sum_{j \neq i}{\gamma^{n+1+i-j}}} 				
\]

\[
\sum_{j \in V}{\gamma^{n+1+i-j}}=\gamma^{n+1}+\sum_{j \neq i}{\gamma^{n+1+i-j}} \Leftrightarrow i \in V			
\]

Where $g \in G$ is a generator of the group $G$ of prime order $p$, and $\gamma$ is picked at random from $Z_p$. 
\subsection{Amendments to CP-WATERS-KEM}
To include Camenisch’s accumulator in the original CP-WATERS-KEM algorithms, the key generation algorithm shall associate an index $i$ to each new generated secret decryption key. The new index $i$ is added to the accumulator. When the Authority needs to revoke a key, it simply removes the corresponding index $i$ and updates the accumulator value. The authority needs also to update the term that is used as base for computing the shared secret (distributed as part of the $MPK$), as it depends on the accumulator.

With the addition or removal of a elements to the accumulator, previously released keys become stale and any party who has previously received a key shall update it. Therefore, a new $\mathtt{UpdateSecretKey}$ step is introduced in the schema. The update algorithm may be executed locally and does not require any secret or computation by the Authority, so to reduce the Authority’s workload.

More in details, the following modifications are needed:

\begin{enumerate}
    \item The authority shall release $acc_V^a$, on behalf of $g^a$; also, in order to be indexed by $i$, the secret decryption key component $K$ shall be modified as follows:  $K=g^{\alpha+abt+b\gamma^i}$ where $b$ is a new private parameter, working as the private parameter $a$ in the original setup algorithm. In fact, $b$ is an uniformly random variable in $Z_p$, kept private by the Authority and $g^b$ shall be released as part of the public key. Furthermore, the Authority shall release $e(acc_V,g)^\alpha e(g^b,g^{\gamma^{n+1}})$ instead of $e(g,g)^\alpha$. This term will be used by the encrypting party to compute the shared secret ${[e(acc_V,g)^\alpha e(g^b,g^{\gamma^{n+1}})]}^s$.
    \item Key generation happens having $L = g^{bt}$ instead of $L = g^t$.
    \item The encrypting party shall release a the ciphertext component $C^\prime$ computed as $C^\prime=g^{bs}$ instead of $C^\prime=g^s$. Each of $C_k$ component shall be computed as $C_k=acc_V^{a\lambda_k}h_{\rho_k}^{-s}$ instead of $C_k=g^{a\lambda_k}h_{\rho_k}^{-s}$. The final ciphertext shall also include a new component: $C^{\prime\prime}=acc_V^s=g^{s \sum_{j \in V}{\gamma^{n+1-j}}}$.
    \item The final equation for decryption is modified as follows:
    \[
    \frac{e(C^{\prime\prime},K)}{\prod_k{[e(C_k,L)e(C^\prime,K_{\rho_k})]^{\omega_k}}e(C^\prime,wit_i)}
    \]
\end{enumerate}

\subsection{CP-WATERS-KEM plus Accumulators} \label{subsec:CP-WATERS-KEM plus Accumulators}
CP-WATERS-KEM is modified as follows:
\begin{enumerate}
    \item $\mathtt{(MPK, MSK) \leftarrow Setup(1^\lambda)}$:  The algorithm outputs the master secret key $MSK$ and the master public key $MPK$, and publishes the $MPK$.
    \begin{itemize}
        \item The algorithm chooses a group $G$ of prime order $p$ and generator $g$, random group elements $h_1, h_2, \dots, h_u$ (where $u$ is the maximum number of system attributes) and a bilinear pairing $e$ such that $e \colon G \times G \to G_T$. In addition, it chooses random exponents $\alpha, a, b, \gamma \in Z_p$.

	\item The algorithm initially creates an accumulator $acc_0=1$, and two initially empty public sets: $V=\{\}$ and $U=\{\}$, where $U$ is the set of all indexes $i$ that will be ever added to the accumulator (but may have been subsequently removed). 

	\item The public key is $g, g^b, h_1, h_2, \dots, h_u,acc_V, acc_V^a$ and 
    \[e(acc_V,g)^\alpha e(g^b,g^{\gamma^{n+1}})\]
    The master secret key is $g^\alpha,\alpha,a,b,\gamma$. 

	\item The sequence $g^\gamma,\dots,g^{\gamma^n} ,g^{\gamma^{n+2}},\dots,g^{\gamma^{2n}}$ (but not $g^{\gamma^{n+1}}$) is made public by the Authority.
    \end{itemize}
    \item $\mathtt{(MPK^\prime,SK_i=(K_i,L_i,\forall_{x \in S} \; K_{i,x},wit_i)) \leftarrow KeyGen(MPK,MSK, S)}$: Key generation happens by taking as input the master keys $(MSK,MPK)$ and a set of attributes $S$ that describe the key. The output is a randomized secret key. The Authority associates an index $i$ to each new generated secret decryption key $SK_i$.
    \begin{itemize}
        \item The algorithm includes $i$ in the set $V$ and $U$: $V=V_{old} \cup \{i\}$, $U=U_{old} \cup \{i\}$ and updates the accumulator $acc_V$:
        \[
        acc_V=\prod_{j \in V}{g^{\gamma^{n+1-j}}}=g^{\frac{\sum_{j \in V}{\gamma^{n+1+i-j}}}{\gamma^i}}
        \]
        
	\item The algorithm updates the terms in the master public key using the accumulator: $acc_V^a$ and $e(acc_V,g)^\alpha e(g^b,g^{\gamma^{n+1}})$, while the master secret key $MSK$ remains the same.

	\item Chosen a random $t \in Z_p$, the algorithm computes the secret decryption key component $K_i$ as as follows: $K_i=g^{\alpha+abt+b\gamma^i}$, $L_i = g^{bt}$  and, for each $x \in S$,  $K_{i,x} = h_x^t$.
 
	\item Also, the algorithm releases a new key component (the witness): 
        \[
        wit_i=\prod_{j \neq i}{g^{\gamma^{n+1+i-j}}}=g^{\sum_{j \neq i}{\gamma^{n+1+i-j}}}.
        \]
    \end{itemize}

    \item $\mathtt{MPK^\prime \leftarrow KeyRemove(PK,MSK, i)}$: A similar step is also executed when the Authority needs to revoke a key $K_i$. In this case, the algorithm simply removes $i$ from the set $V$, recomputes the accumulator value $acc_V$ and, consequently, the terms in the master public key using it:  $acc_V^a$ and $e(acc_V,g)^\alpha e(g^b,g^{\gamma^{n+1}})$ as above (the master secret key $MSK$ remains the same).
    
    With the addition or removal of elements to the accumulator, previously released witnesses become stale and any Client who has previously received a witness $wit_i$ shall update it. Therefore, the following step is introduced into the schema:

    \item $\mathtt{wit_i^\prime \leftarrow UpdateWitness(MPK, V_{old}, V, wit_i})$: The algorithm takes as input the old witness and updates it to match the new master public key $MPK$ and the current set of authorized indices $V$. 
    The new witness is locally computed using the following equation:
    \[
    wit_{i}^\prime \leftarrow {wit_i} \frac{\prod_{j \in V/V_{old}}{g^{\gamma^{n+1+i-j}}}}{\prod_{j \in V_{old}/V}{g^{\gamma^{n+1+i-j}}}}
    \]
    Note that the Client does not know $g^{\gamma^{n+1}}$, hence, this algorithm fails when the condition $i \in V \cap V_{old}$ is not verified, i.e. a Client cannot update its $wit_i$ if $i$ is not (no more) in $V$ as a result of a revocation. In this case the update operation returns $wit_i^\prime=\bot$.

    \item $\mathtt{(C=(C^\prime,C^{\prime\prime}, \forall_{k \in [1,\dots\,l]} \; C_k), \mu)\leftarrow Encrypt(MPK, M^{l \times m}, \rho)}$: The algorithm takes as input an access structure $(M, \rho)$ and the public key $MPK$. $M$ is an $l \times n$ matrix, while $\rho$ is an injective function associating each row of $M$ to an attribute $\rho_k$ (i.e.,$\rho_k=\rho(k) \in S$); note that in this construct one attribute is associated with at most one row. The output is a random secret and the ciphertext.
    \begin{itemize}
        \item 	Chosen a random vector $\vec{v}=(s,y_2,\dots,y_m)$ in $Z_p^n$ and being $M_k$ the $k$-th row of $M$, the algorithm computes $\lambda_k=v\cdot M_k$.

	\item Together with a with description of $(M, \rho)$, the algorithm makes public the ciphertext: 
    \[
    C^\prime=g^{bs}, C_k= acc_V^{a\lambda_k}h_{\rho_k}^{-s},
    C^{\prime\prime}=acc_V^s={(\prod_{j \in V}{g^{\gamma^{n+1-j}}})}^s
    \]
    \item Finally, the algorithm computes the random secret 
    \[
    {\mu=[e(acc_V,g)^\alpha e(g^b,g^{\gamma^{n+1}})]}^s
    \]
    and keeps it private.
    \end{itemize}
    \item $\mathtt{\mu \leftarrow Decrypt(SK_i, C)}$: Dually, the decryption takes as input the ciphertext $C$ and the secret key $SK$. The output is the shared secret if and only if the set of attributes $S$ satisfies the access structure, or $\bot$ otherwise.
    \begin{itemize}
        \item For each $k$ such that $\rho_k \in S$ (i.e., consider only attributes in $S$), compute $\omega_k$ such that $\sum_k{\omega_k\lambda_k=s}$ (there could different sets of  $\{ \omega_k \}$ satisfying this equation)
        \item Compute the random secret:
        \[\mu=
         \frac{e(C^{\prime\prime},K_i)}{\prod_k{[e(C_k,L_i)e(C^\prime,K_{i,\rho_k})]^{\omega_k}}e(C^\prime,wit_i)}=\]
         \[{[e(acc_V,g)^\alpha e(g^b,g^{\gamma^{n+1}})]}^s
        \]
    \end{itemize}
\end{enumerate}
\subsection{Correctness}
To understand why decryption works, consider the solution equation, and note the numerator is
\[
e(C^{\prime\prime},K)=
e(g,g)^{\frac{\sum_{j \in V}{\gamma^{n+1+i-j}}}{\gamma^i}s(\alpha+abt+b\gamma^i)}=
\]
\[
e(g,g)^{\frac{\sum_{j \in V}{\gamma^{n+1+i-j}}}{\gamma^i}\alpha s}
e(g,g)^{\frac{\sum_{j \in V}{\gamma^{n+1+i-j}}}{\gamma^i}sabt}
e(g,g)^{bs{\sum_{j \in V}{\gamma^{n+1+i-j}}}}=
\]
\[
e(acc_V,g)^{\alpha s}
e(acc_V,g)^{sabt}
e(g^s,g)^{b{\sum_{j \in V}{\gamma^{n+1+i-j}}}}
\]

As in the original CP-ABKEM decryption algorithm, the second factor $e(acc_V,g)^{sabt}$ cancels out with the first part of the denominator:
\[
\prod_k[e(C_k,L)e(C^\prime,K_{\rho_k})]^{\omega_k}=
\prod_k{e([acc_V^{a\lambda_k}h_{\rho_k}^{-s},g^{bt})e(g^{bs},h_{\rho_k}^t)]^{\omega_k}}=
\]
\[
\prod_k{e(acc_V,g)^{ab\lambda_k \omega_k t}}=
e(acc_V,g)^{sabt}
\]

Regarding the third factor $e(g^s,g)^{b{\sum_{j \in V}{\gamma^{n+1+i-j}}}}$ we note that, if and only if the index is contained in the current accumulator (i.e., $i \in V$), we have:

\[
e(g^s,g)^{b{\sum_{j \in V}{\gamma^{n+1+i-j}}}}=
e(g^{bs},g^{\gamma^{n+1}+\sum_{j \neq i}{\gamma^{n+1+i-j}}})=
\]
\[
e(g^{bs},g^{\gamma^{n+1}})
e(g^{bs},wit_i)
\]

Partially cancelling out with the factor $e(g^{bs},wit_i)$ in the denominator. 
Therefore, the result of the computation is \[{[e(acc_V,g)^\alpha e(g^b,g^{\gamma^{n+1}})]}^s\].

\subsection{Security} \label{subsec:Security}
To prove security, we use a security game based on the one presented in Section 5 of \cite{cryptoeprint:2008/290}. The adversary chooses to be challenged on an encryption to an access structure $A^*$, and can ask arbitrarily $q$ times for any private key $S$ that does not satisfy $A^*$. However, the original model is extended by letting the adversary query for private keys that satisfy the access structure, with the restriction that any of those keys shall be revoked before the challenge:

\begin{itemize}
    \item \emph{Setup}. The challenger runs $\mathtt{Setup}$ algorithm and gives the public parameters, $\mathtt{PK}$ to the adversary.
    \item \emph{Phase 1}. The adversary makes repeated private keys corresponding to sets of attributes $\mathtt{S_1, \dots , S_{q^\prime}}$ (with $1<q^\prime<q$). 
    \item \emph{Revocation} Using the $\mathtt{keyremove}$ algorithm in the schema, any key may (or not) be revoked.
    \emph{Phase 1} and \emph{Revocation} may be arbitrarily interleaved.
    \item \emph{Challenge}. The adversary submits two equal length messages $\mathtt{M_0}$ and $\mathtt{M_1}$. In addition the adversary gives a challenge access structure $\mathtt{A^*}$ such that none of the sets $\mathtt{S_1, \dots , S_{q^\prime}}$ from \emph{Phase 1} satisfies the access structure, or such that any of those keys corresponding to sets satisfying the access structure has been revoked. The challenger flips a random coin $\mathtt{\beta}$, and encrypts $\mathtt{M_\beta}$ under $\mathtt{A^*}$. The ciphertext $\mathtt{CT^*}$ is given to the adversary. 
    \item \emph{Phase 2}. \emph{Phase 1} is repeated with the restriction that none of sets of attributes $\mathtt{S_{q^\prime+1},\dots,S_q}$ satisfies the access structure corresponding to the challenge. Revocation may also occur in this phase.
    \item \emph{Guess}. The adversary outputs a guess $\mathtt{\beta^\prime}$ of $\mathtt{\beta}$. 
\end{itemize}
The advantage of the adversary in the above game is $\epsilon=\mathtt{Pr[\beta^\prime=\beta]-\frac{1}{2}}$ and, by definition, the scheme is secure if all polynomial time adversaries have at most a negligible advantage.
We use a selective proof, therefore the above game is augmented by an initial step \emph{Init} in which the adversary commits to the challenge access structure $A^*$ and to the final set of credentials that will eventually appear in the accumulator $V^*$.

Our security proof works under the General Diffie-Hellman Exponent Problem introduced by Boneh, Boyen and Goh in \cite{cryptoeprint:2005/015}. Let $\nu$ and $\sigma$ be natural numbers, $P, Q \in Z^+_p[X_1, \dots,Xn]^\sigma$ be two s-tuples (i.e., ordered sets) of n-variate polynomials over $Z^+_p$ and let $f \in Z^+_p[X_1,\dots,X_\nu]$. The notation $P = (p_1, p_2,\dots, p_\sigma), Q = (q_1, q_2,\dots, q_\sigma)$ is used to refer to each set of polynomials and it is required that $p_1 = q_1 = 1$; $f$ is said dependent on $P$ and $Q$ if there exist $\sigma^2+\sigma$ constants $\{a\}_{i,j=1}^\sigma, \{b_k\}^\sigma_{k=1}$ such that 
\[ f^*=\sum{a_{i,j}p_ip_j}_{i,j=1}^\sigma+\sum{b_k}^\sigma_{k=1} \]
$f^*$ is independent of $P$ and $Q$  iff is not dependent on $P$ and $Q$. Furthermore, consider a bilinear map $e:G_0 \times G_0 \to G_1$ and two injective maps $\xi_0$, $\xi_1$ of the additive group $Z^+_p$ , i.e.  $\xi_0, \xi_1: Z^+_p \to \{0, 1\}^m$ such that $G_i=\{\xi_i(x) : x \in Z^+_p \}$, for $i=0,1$. Use  $deg(f)$ to denote the degree of any function $f$ and define $deg(P)= max\{deg(f) : f \in P\}$. The Complexity Lower Bound theorem in Generic Bilinear Groups theorem states that,  if
$f^*$ is independent of $P$ and $Q$ then, the advantage of any adversary $A$ that makes a total of at most $q$ queries to an oracle computing group operations in $G_0,G_1$ and the bilinear pairing $e:G_0 \times G_0 \to G_1$ in distinguish $f^*(x_1,\dots,x_n)$ from a random group element in $Z_p$ cannot exceed $\frac{(q+2\sigma+2)^2\cdot d}{2p}$, where $d = max(2\cdot deg(P) , deg(Q), deg(f)).$
Consequently, the General (Decisional) Diffie-Hellman Exponent Problem, which is formulated as follows: given $g_0 \in G_0, g_1 \in G_1$,
\[
(g_0^{P(x_1,\dots,x_n)},g_1^{Q(x_1,\dots,x_n)}) \in G_0^s \times G_1^s
\]
distinguish $g_1^{f^*(x_1,\dots,x_n)} \in G_1$ from a random element in $G_1$, is a hard problem in the generic group model.

Using the aforementioned hardness assumption, in Appendix \ref{appendix:1} we prove that chosen an access structure $A^*$, no polynomial time adversary can (selectively) break our system, provided all keys satisfying $A^*$ have been revoked before the challenge.

Note that the presented security model supports only chosen-plaintext attacks. The model is extended to handle chosen-ciphertext attacks by allowing for decryption queries in Phase 1 and Phase 2. To achieve chosen-ciphertext security we use the Fujisaki-Okamoto transformation reported in subsection \ref{subsec:CCA-secure Encryption Algorithm} and \ref{subsec:CCA-secure Decryption Algorithm}. As this transformation exactly applies as in the original paper, we let the reader refer to \cite{OABE} for the security proof. 

\subsection{Verification Protocol}
We use the above CP-ABE schema to implement the revocable functional credential schema in section \ref{subsec:Revocable Functional Credentials}(the protocol is also adapted from \cite{TS103532}):
\begin{enumerate}
    \item CP-ABE $\mathtt{Setup(1^\lambda)}$ algorithm takes place in order to generate the key pair $(MPK,MSK)$. 
    \item The grant credential algorithm is implemented through the $\mathtt{MPK^\prime,SK_i \leftarrow KeyGen(MPK,MSK, S)}$ algorithm which releases credentials $cred_i= SK_i$ corresponding to a non-empty set of attributes $S$. It also outputs an updated master public key $MPK^\prime$. 
    \item To check credential, chosen an access policy (i.e., a matrix $M^{l \times m}$, and a function $\rho$), a verifier generates and encrypts a random secret $\mu$ through the CP-ABE $\mathtt{Encrypt(MPK, M^{l \times m}, \rho)}$; and sends the resulting ciphertext $C$ to the prover. Using a credential $SK^\prime$, the prover executes $\mathtt{\mu^\prime \leftarrow Decrypt(SK^\prime, C)}$ and sent back the result to the verifier. The verifier output $1$ if $\mu=\mu^\prime$ and $0$ otherwise.
    \item Revocation of credential $cred_i$ is implemented through the $\mathtt{KeyRemove(PK,MSK, i)}$ algorithm, which updates the master public key to $MPK^\prime$.
\end{enumerate}
Note that since each challenge encapsulates a randomly-generated secret token, the protocol is natively immune to reply attacks.
It is easy to prove that the protocol satisfies the correctness and unforgeability properties as above defined for functional credentials. Section \ref{sec:Anonimity} provides a proof of the anonymity property. 

\section{Anonymity} \label{sec:Anonimity}
To ensure anonymity, the Fujisaki Okamoto transformation may be applied to CP-WATERS-KEM. This transformation was already described in \cite{OABE} and there proved to make the schema secure under Chosen Ciphertext Attacks. The transformation uses two random numbers $r_{c}$ and $K_{c}$ both chosen by the encrypting party to generate randomness for encryption. It is possible to prove that there is a negligible probability for an attacker to produce a ciphertext that may decrypt, and that this probability is the same as guessing a ciphertext without any knowledge of the randomness used to produce it (see \cite{OABE}, Lemma 3.1.5). 

\subsection{CCA-secure Encryption Algorithm} \label{subsec:CCA-secure Encryption Algorithm}
The CCA-secure encryption algorithm is specified by the following steps:
\begin{itemize}
    \item The decrypting party (prover) shall choose a random number $r_{c}$ and send it to the encrypting party.
    \item Received $r_{c}$, the encrypting party (verifier) chooses an access structure $AP$ and a secret $K_{c}$ and concatenates them to form the string $r_{c}||K_{c}||AP$
    \item The encrypting party runs the encryption algorithm of the original CP-WATERS-KEM or of the modified schema with revocation to get a random secret and the ciphertext. The seed $r_{c}||K_{c}||AP$ is used as a source of randomness for the encryption algorithm with $\mathtt{u \leftarrow PRG(H^\prime(r_{c}||K_{c}||AP),\lambda)}$, where $\mathtt{PRG}$ is a pseudo random generator, $\lambda$ is the length of the returned random bit string ($u \in \{0,1\}^l$) and $H^\prime$ is a collision-resistant hash function.
    \begin{itemize}
        \item The random secret is $e(g,g)^{\alpha s}$ for CP-WATERS-KEM or $(e(acc_V,g)^\alpha e(g^b,g^{\gamma^{n+1}}))^s$ for the modified CP-WATERS-KEM. The encrypting party keeps it private and uses in the next step.
        \item The encrypting party releases the ABKEM ciphertext $C_{ABKEM}$.
    \end{itemize}
    \item The encrypting party uses random secret above for XORing the concatenation $r_{c}||K_{c}$
        \begin{itemize}
            \item Transform $r_{c}||K_{c}$ into bytes (octects).
	    \item Using the pseudo random generator $PRG$, get 
     \[\mathtt{r \leftarrow PRG(H(e(g,g)^{\alpha s}),\lambda)}\] 
     for CP-WATERS-KEM or 
     \[\mathtt{r \leftarrow PRG(H([e(acc_V,g)^\alpha e(g^b,g^{\gamma^{n+1}})]^s),\lambda)}\]
     for the modified CP-WATERS-KEM, with $H$ being a collision-resistant hash function.
            \item Finally, compute $C = r \oplus (K_{c}||r_{c})$
        \end{itemize}
\end{itemize}

\subsection{CCA-secure Decryption Algorithm} \label{subsec:CCA-secure Decryption Algorithm}
The CCA-secure decryption algorithm is specified by the following steps:
    \begin{itemize}
        \item Run decryption of the original CP-WATERS-KEM or of the modified schema to decrypt the ciphertext and obtain the shared secret: 
        \[e(g,g)^{\alpha s}\] for CP-WATERS-KEM or 
        \[(e(acc_V,g)^\alpha e(g^b,g^{\gamma^{n+1}}))^s\] for the modified schema.
        \item Use that shared secret to generate randomness \[\mathtt{r \leftarrow PRG(H(e(g,g)^{\alpha s}),\lambda)}\] or \[\mathtt{r\leftarrow PRG(H([e(acc_V,g)^\alpha e(g^b,g^{\gamma^{n+1}})]^s),\lambda)}\] 
	\item Use generated randomness $r$ for XORing the ciphertext and retrieve $K_{c}$ and $r_{c}$: $C \oplus r = (K_{c}||r_{c})$
	\item Verify $r_{c}$ matches the random number chosen at beginning.
	\item Run again the CCA-secure Encryption using $r_{c}||K_{c}||AP$ as a source of randomness and verify the result is equal to the received ciphertext $C_{ABKEM}$.
    \end{itemize}

\subsection{Security}
We prove the following theorem:
\begin{theorem}A polynomial time adversary, acting as a Verifier, cannot distinguish between any two provers with different CP-WATERS-KEM keys, if their keys both satisfy (or not satisfy) the same access structure they are tested against. 
\end{theorem}
\begin{proof}
We start considering the following security game (adapted from \cite{FCREDENTIALS}):
\begin{enumerate}
    \item The Setup algorithm of CP-WATERS-KEM or the modified schema takes place. The public key $PK$ is given to the adversary. 
    \item Any Prover $P_i$ receives distinct secret keys $K_i$ embedding some attributes. 
    \item \label{step:3} The adversary is allowed to submit queries in the form $(r_c||K_c||AP)$ to an oracle which produces a random output $u$ if this is the first time the input has been queried on. Otherwise, it gives back the previous response. In addition, the oracle computes the ciphertext $C$ using the CCA-secure encryption algorithm and records the couple $((r_c||K_c||AP),(C,u))$ in a table. This oracle operation is run throughout the whole game.
    \item The adversary, acting as a Verifier $V$, arbitrarily chooses an access structure $A^*$ and two Provers $P_0$ and $P_1$, such that their corresponding keys either both satisfy, or both not satisfy the chosen access structure.
    \item Depending on an internal coin toss $b$, an oracle impersonates prover $P_b$ in the verification algorithm. 
    \item Verifier $V$ computes a CCA-secure ciphertext and sends it to the oracle. 
    \item The oracle responds with the decrypted ciphertext $m$ or with $\bot$. 
    \item The aforementioned steps (except the Setup) are repeated adaptively for any polynomial number of times on arbitrarily chosen access structure and arbitrarily chosen pairs of provers. 
    \item The Verifier try a guess $b^\prime$ and wins the game if $b==b^\prime$ (i.e., she is able to guess which Prover has responded). 
\end{enumerate}

Modify the game as follows: at step 7, when given a ciphertext $C$, the oracle checks if $C$ appears in the random oracle table. If so, it outputs the corresponding $m=(K_c||r_c)$ value in the table; otherwise, it outputs $\bot$ and rejects.

The difference between the original game and the modified one is negligible, as in the original game the oracle may decrypt even in case of a forged ciphertext (i.e., a ciphertext not computed using the CCA-secure encryption algorithm). However, since the oracle was not queries on $(r_c||K_c||AP)$, the probability that this event happens is bounded by the probability of apriori guessing a ciphertext output by an encryption for a given message without knowing the randomness used to encrypt.

Now the following observations apply to this modified game:
\begin{itemize}
    \item If the Verifier produces a genuine ciphertext C following the CCA-secure Encryption algorithm, she gets a correct decryption $m$ if the attributes embedded in the secret key $K_b$ satisfy the chosen access structure $A^*$, i.e. $A^*(K_b)=1$. Thus, the presented schema satisfies by definition the correctness property. 

    \item Viceversa, if the attributes embedded in the secret key $K_b$ do not satisfy the chosen access structure $A^*$, i.e. $A^*(K_b)=0$, the ciphertext wouldn’t decrypt at all except for a negligible probability $\epsilon$. Thus, the presented schema satisfies  by definition the unforgeability property.
\end{itemize}

Furthermore, we observe that: 

\begin{itemize}
    \item The access structure $A^*$ associated to the ciphertext $C$ is always known to the challenger (given as input after being chosen by the adversary)
    \item Because a pseudo random generator is used, the ciphertext $C$ is deterministically computed from the public key $PK$ and the access structure $A^*$
    \item The ciphertext $C$ is uniformly distributed on the ciphertext space, because computed using the uniformly distributed randomness $u$ in step \ref{step:3}.
    \item No decryption happens when the Verifier produces a forged ciphertext.
\end{itemize}

Under the conditions above, suppose to modify the previous game replacing prover $P_b$’s behaviour as follows: 
\begin{itemize}
    \item if key $K_b$ embeds attributes satisfying the access structure $A^*$, then message $m$ is returned;
    \item otherwise $\bot$ is returned. 
\end{itemize}

That is, $P_b$ no longer evaluates the decryption using the key $K_b$ rather it (deterministically) returns $m$ or $\bot$ depending on the internal bit $A^*(K_b)$. Since $A^*(K_0)$ = $A^*(K_1)$ (both keys satisfy or not satisfy the access structure), in the latter schema the random coin $b$ of the oracle remains hidden in an information-theoretic sense. This implies that the advantage of verifier $V$ is $1/2$ in distinguish between $P_0$ and $P_1$. As the introduced modification does not alter the advantage of the verifier $V$ except for at most a negligible probability $\epsilon$, the advantage of $V$ in the original schema is negligibly close to $1/2$. $\square$
\end{proof}

\section{Related Works}
A survey on revocation strategies for anonymous credentials is presented in \cite{10.1007/978-3-642-24712-5_1}. Early attempts included time-based attributes contained in the credentials or reissuing of credentials after a change in the key material. Similar approaches work as well for functional credentials using predicate encryption. However they may be unpractical in real-world contexts.
Whitelists and blacklists approaches may be complementary used as well. To invalidate revoked credentials, the issuer shall periodically update the white or black list. For predicate encryption, a trivial naive approach would be to implement directly  whitelists or blacklists in the access policy (\emph{verifier-local revocation}). However, this would typically increment the length of the policy linearly with the number of involved credentials. More dangerously, this approach leaves the control on revocation to the verifier, not to the issuer. This is formally incorrect and may potentially expose the prover to risk of \emph{backward linkability} (a similar problem was highligthed in group signature schemas \cite{10.1145/501983.502015}). For functional credentials in particular, the verifier can re-identify the prover by specially crafted policies able to create correlation when the same credential is shown more than once.

Building on Dual-Policy Attribute-Based Encryption (DP-ABE), \cite{ambrona2017generic} presents a predicate encryption scheme that supports revocation by combining Boolean formula predicate encoding with broadcast encryption. The latter encoding takes care of revocation, while the former encodes the desired access structure. A drawback is that the setup, key generation, and encryption time grows polynomial with the number of users.

\cite{10.1007/978-3-319-27239-9_4} reports several works considering the application of dynamic universal accumulators to anonymous credentials to implement blacklists. While these approaches require to prove in zero-knowledge that a prover's non-membership witness satisfies the accumulator verification equation, the authors describe a different construction where both the accumulator and the anonymous credentials, previously described in \cite{10.1007/978-3-662-45611-8_26}, rely on the same construct (structure-preserving signatures on equivalence classes). To same extent, our approach is similar to their one, but we highlight the different scope as \cite{10.1007/978-3-319-27239-9_4} limits to consider \emph{selective disclosure}, not proof of predicates. In terms of performance, scheme 2 in \cite{10.1007/978-3-662-45611-8_26}, using primitives $\mathtt{VerifyR}$ and $\mathtt{VerifySubset}$, requires a total of $2*(i+2)$ pairing operations per number of credential entries $i$, plus two additional revocation-induced pairings (scheme 2 in \cite{10.1007/978-3-319-27239-9_4}). Our scheme presented in section \ref{subsec:CP-WATERS-KEM plus Accumulators}, using the optimization described in \cite{OABE}, reduces the number of pairings to $k + 2$, with $k$ being the number of attributes satisfying the policy, plus one more for checking the witness.

\section{Conclusion}
Combining Ciphertext Policy Attribute-Based Encryption (CP-ABE) and accumulators we build a revocable credential management framework supporting anonymous proof of predicates over attributes; we further achieve anonymity by applying a simple transformation to the resulting schema. To the best of our knowledge, our work is the first one efficiently combining rich policy expressiveness (from ABE), revocation (from accumulator) and anonymous proof of predicates over attributes into a single framework.
\bibliographystyle{splncs04}
\bibliography{bibliography}
\appendix
\section{Proof of Theorem in Section \ref{subsec:Security}} \label{appendix:1}
\begin{theorem}. Under the General Decisional Diffie-Hellman Exponent hardness assumption (\cite{cryptoeprint:2005/015}), chosen an access structure $A^*$ and a set of non revoked credentials $V^*$, no polynomial time adversary can selectively break our system, provided all credentials satisfying $A^*$ are not in $V^*$. 
\end{theorem}
\begin{proof} Consider an adversary $A$ with non-negligible advantage $\epsilon=Adv_A$ that plays a security game against our construction.

\begin{itemize}
    \item \emph{Init}. The adversary chooses $A^*=(M^*,\rho^*)$ and $V^*$ (i.e. the set of keys not satisfying the access structure that will eventually stay in $V$ just before the challenge) and gives it to the challenger. 
    \item \emph{Setup}. Given a group $G$ of prime order $p$ and generator $g$ and a bilinear pairing $e$ such that $e \colon G \times G \to G_T$, the simulator takes in: 
    \begin{itemize}
        \item the terms $g^\alpha, g^a, g^b, g^{ab}, g^{\alpha s^* }, g^{bs^*}$
        \item for each $i \in V^*$, the terms $g^{s^* \gamma^{n+1-i}}$
	\item for each $i$ the sequence $g^\frac{b \gamma^i}{a},g^\frac{b \gamma^i}{a^2},\dots,g^\frac{b \gamma^i}{a^{m^*}}$, but not $g^{b \gamma ^i}$
	\item the sequence $g,g^\gamma,\dots,g^{\gamma^n},g^{\gamma^{n+2}}, \dots,g^{\gamma^{2n}}$, but not $g^{\gamma^{n+1}}$
        \item the sequence $g,g^{a^j \gamma},\dots,g^{a^j \gamma^n},g^{a^j \gamma^{n+1}},\dots, ,g^{a^j \gamma^{2n}}$ – including $g^{a^j \gamma^{n+1}}$, for each $a^j$ with $j \in [-m^*, m^*-1]/\{0\}$
        \item $T$ which is either a random element in $G_T$ or $e(g,g)^{bs^* \gamma^{n+1}}$. 
    \end{itemize}

    Note that from the above given terms, the simulator is not able to compute $g^{\gamma^{n+1}}$ in polynomial time and therefore cannot directly compute $e(g,g)^{bs^* \gamma^{n+1}}$ as $e(g^{bs^*},g^{\gamma^{n+1}})$.
    The challenger chooses random group elements $h_1, h_2, \dots, h_u$ where $u$ is the maximum number of system attributes. 
    
    The simulator creates an empty accumulator $acc_V=acc_0=1$, and two empty public sets: $V=\{\}$ and $U=\{\}$.
    
    Chosen $z_x \in Z_p$, for each $x$, if $\rho^*(k)=x$, $h_x$ is computed as 
        \[
        h_x= acc_{V^*}^{z_x} \cdot acc_{V^*}^{aM^*_{k,1}} \cdot acc_{V^*}^{a^2 M^*_{k,2}} \cdot \dots \cdot acc_{V^*}^{a^{m^*} M^*_{k,{m^*}}} =
        \]
        \[
        (\prod_{j \in V^*}{g^{\gamma^{n+1-j}}})^{z_x} 
        \prod_{l=1}^{m^*}{(\prod_{j \in V^*}{g^{a^l \gamma^{n+1-j}}})^{M^*_{k,l}}}  
        \]
        otherwise let $h_x= acc_{V^*}^{z_x}$
    
    The challenger releases the public key $MPK$: 
        \[g, g^b, h_1, h_2, \dots, h_u\]
        \[acc_V=\prod_{j \in V}{g^{\gamma^{n+1-j}}}\]
        \[acc_V^a=\prod_{j \in V}{g^{a \gamma^{n+1-j}}}\] and the sequence $g^{\gamma},\dots,g^{\gamma^n},g^{\gamma^{n+2}},\dots,g^{\gamma^{2n}}$ (but not $g^{\gamma^{n+1}}$).
        
    The challenger also releases 
        \[
        e(acc_V,g)^\alpha e(g,g)^{b \gamma^{n+1}}=e(acc_V,g^\alpha) e(g^{\frac{b}{a^j}\gamma^i},g^{a^j\gamma^{n+1-i}})
        \]
    
    \item \emph{Phase 1}. The adversary makes repeated private key queries corresponding to sets of attributes that either not satisfy the access structure or satisfy it but will not be eventually in $V^*$:
        \begin{itemize}
            \item chosen $r_i \in Z_p$, $t_i=r_i+(w_1 a^{-1}+w_2 a^{-2}+\dots+w_{m^*} a^{-m^*})\gamma^i$ where $\vec{w}=(w_1,\dots,w_{m^*}) \in Z_p^{m^*}$ and $w_1=-1$:
            
            \[K_i=g^{\alpha+ab{t_i}+b\gamma^i}=\]
            \[g^{\alpha+ab{r_i}-b\gamma^i+\frac{w_2}{a}b\gamma^i+\frac{w_3}{a^2}b\gamma^i+\dots+\frac{w_{m^*}}{a^{m^*-1}} b\gamma^i+b\gamma^i}=\]
            \[
            g^{\alpha+ab{r_i}}(\prod_{j=2}^{m^*}{g^\frac{b\gamma^i}{a^{j-1}}})^{w_j}
            \]
            \[
            L_i=(g^{t_i})^b=\]
            \[g^{b{r_i}+b(w_1 a^{-1}+w_2 a^{-2}+\dots+w_{m^*} a^{-m^*})\gamma^i}=\]
            \[g^{b{r_i}} \prod_{j=1}^{m^*}{(g^\frac{b\gamma^i}{a^j})^{w_j}} 
            \]
            Note that the factor $g^{b\gamma^i}$  is not needed to compute $K_i$ nor $L_i$. 
            \item For each $x_k \in S_i$ where $\rho^*(k)=x_k$, $K_{i,x_k}$ is computed as:
            \[
            K_{i,x_k}=h_{x_k}^{t_i}=\]
            \[(acc_{V^*}^{z_x} \cdot acc_{V^*}^{aM^*_{k,1}} \cdot acc_{V^*}^{a^2 M^*_{k,2}} \cdot\]
            \[\dots \cdot acc_{V^*}^{a^{m^*} M^*_{k,{m^*}}})^{r_i+(w_1 a^{-1}+w_2 a^{-2}+\dots+w_{m^*} a^{-m^*})\gamma^i}=
            \]
            \[
            acc_{V^*}^{(z_x+aM^*_{k,1}+a^2 M^*_{k,2}+\dots+a^{m^*} M^*_{k,{m^*}})(r_i+(w_1 a^{-1}+w_2 a^{-2}+\dots+w_{m^*} a^{-m^*})\gamma^i)}=
            \]
            \[
            (\prod_{j \in V^*}{g^{\gamma^{n+1-j}}})^{r_i z_x}
            \prod_{l=1}^{m^*}{(\prod_{j \in V^*}{g^{a^l \gamma^{n+1-j}})^{r_i M^*_{k,l}}}} \cdot\]
            \[
            \prod_{l=1}^{m^*}{(\prod_{j \in V^*}{g^\frac{(\gamma^{n+1-j}) \gamma^i}{a^l})^{w_l z_x}}} 
            \prod_{o=1}^{m^*}{\prod_{l \neq o}^{m^*}{(\prod_{j \in V^*}{g^{a^{o-l} (\gamma^{n+1-j}) \gamma^i})}^{w_l M^*_{k,o}}}} \cdot\]
            \[
            \prod_{l=1}^{m^*}{(\prod_{j \in V^*}{g^{\gamma^{n+1-j} \gamma^i}})^{w_l M^*_{k,l}}}
            \]
            The last term apparently requires $g^{\gamma^{n+1}}$ when $i \in V^*$. However, we note that for those keys not satisfying the access structure, for any row $k$ of $M^*$ where $\rho^*(k)=x_k$ (i.e., the attribute $x_k \in S_i$ is used in the access structure) we can choose $w_2  \dots  w_{m^*}$ so that the equation $\vec{w}M^*_k=0$ holds, and
            \[
            \prod_{l=1}^{m^*}{(\prod_{j \in V^*}{g^{\gamma^{n+1-j} \gamma^i}})^{w_l M^*_{k,l}}} = 1
            \]

           Therefore, computing $K_{i,x_k}$, for each $i \in V^*$ not satisfying the access structure does need terms like $g^{a^l (\gamma^{n+1-i}) \gamma^i }=g^{a^l \gamma^{n+1}},l \neq 0$, but never uses $g^{\gamma^{n+1}}$. For those keys satisfying the access structure (where $\vec{w}M^*_k=0$ does not hold), but not in $V^*$ we note that in the above term $\gamma^{n+1-j} \gamma^i \neq \gamma^{n+1}$. Therefore, we do not use $g^{\gamma^{n+1}}$ to compute $K_{i,x_k}$.
           \item For each $x_k \in S_i$ where $x_k$ is not used in the access structure, $K_{i,x_k}$ is computed as:
            \[
            K_{i,x_k}=h_{x_k}^{t_i}=\]
            \[
            acc_{V^*}^{z_x (r_i+w_1 a^{-1}+w_2 a^{-2}+\dots+w_{m^*} a^{-m^*})}=\]
            \[
            (\prod_{j \in V^*}{g^{\gamma^{n+1-j}}})^{z_x r_i} 
            \prod_{l=1}^{m^*}{(\prod_{j \in V^*}{g^\frac{\gamma^{n+1-j}}{a^l}})^{w_l}}
            \]

            \item The witness is computed as: $wit_i=\prod_{j \neq i}{g^{\gamma^{n+1+i-j}}}$
            \item The simulator includes $i$ in the set $V$ and $U$: $V=V_{old} \cup \{i\}, U=U_{old} \cup \{i\}$ and updates the accumulator $acc_V=\prod_{j \in V}{g^{\gamma^{n+1-j}}}$
            \item Consequently, the simulator updates the terms in the public key using the accumulator: $acc_V^a=\prod_{j \in V}{^{a \gamma^{n+1-j}}}$ and computes
            \[
            e(acc_V,g)^\alpha e(g^b,g^{\gamma^{n+1}})=e(acc_V,g)^\alpha e(g^{\frac{b}{a^j}\gamma^i},g^{a^j\gamma^{n+1-i}})
            \]
        \end{itemize}
    \item \emph{Revocation}. The simulator revokes any key $K_i$ which is asked by removing $i$ from set $V$. Each key owner updates her witnesses as usual (except for witnesses of removed keys). The public parameters are updated as usual. 
\emph{Phase 1} and \emph{Revocation} are repeated (even interleaved each other) so that any key satisfying the access structure is removed. Hence, $V^*$ is the final set of keys in the accumulator.
    \item \emph{Challenge}. Finally the adversary submits two equal length messages $M_0$ and $M_1$. The challenger flips a random coin $\beta$, and encrypts $M_\beta$. The challenger gives to the adversary the ciphertext:
    \[
        \{M_\beta\}_{A^*}=M_\beta (e(acc_V,g)^\alpha e(g^b,g^{\gamma^{n+1}}))^{s^*}=\]
        \[
        M_\beta (e(acc_V,g^{\alpha s^*})T)
    \] 
    and the terms:
    \[
        C^{\prime *}=g^{b s^*}
    \]
    \[
        C^{\prime\prime *}=acc_{V^*}^{s^*}=
        \prod_{j \in V^*}{g^{s^*\gamma^{n+1-j}}}  
    \]

    Chosen $\vec{v}=(s^*,s^* a+y^\prime_2,\dots,s^* a^{m^*-1}+y^\prime_{m^*})$ and $\lambda_k=M^*_k \vec{v} = M^*_{k,1} s^*+M^*_{k,2} (s^* a +y^\prime_2)+\dots+M^*_{k,m^*} (s^* a^{m^*-1}+y^\prime_{m^*})$ and given
    \[
    h_{\rho_k}^{-s^*}=\]
    \[(acc_{V^*}^{s^*})^{-z_{\rho_k}} 
    acc_{V^*}^{a s^* M^*_{k,1}} \cdot
    acc_{V^*}^{a^2 s^* M^*_{k,2}} \cdot
    \dots \cdot
    acc_{V^*}^{a^{m^*} s^* M^*_{k,m^*}}
    \]
    The term $C^*_k$  can be finally computed as
    \[
    C^*_k=acc_{V^*}^{a \lambda_k} h_{\rho_k}^{-s^*}=\]
    \[
    (acc_{V^*}^{s^*})^{-z_{\rho_k}} \prod_{l=2}^{m^*}{(acc_{V^*})^{a M^*_{k,l} y^\prime_l}}=\]
    \[
    \frac{\prod_{l=2}^{m^*}{(\prod_{j \in V^*}{g^{a \gamma^{n+1-j}})^{M^*_{k,l} y^\prime_l}}}}{(\prod_{j \in V^*}{g^{s^* \gamma^{n+1-j}})^{z_{\rho_k}}}}
    \]
    \item \emph{Phase 2}. Phase 1 is repeated with the restriction that none of the sets of attributes for which a key is requested satisfies the access structure $A^*$. Revocation may also occur in this phase.
    \item \emph{Guess}. After querying for at most $q$ keys, the adversary outputs a guess $\beta^\prime$ of $\beta$ with probability $\epsilon=Pr[A(\beta^\prime==\beta)]$. When $T$ is a tuple, the adversary guesses  $e(g,g)^{b s^* \gamma^{n+1}}$ with advantage $\epsilon$. 
    \end{itemize}
Using the terminology of the generic proof template by Boneh, Boyen, and Goh (introduced in Section \ref{subsec:Security}), we observe that the monomial $f^*=b s^* \gamma^{n+1}$ is independent of the two sets of terms
    \[
    Q=\{1\}
    \]
    \[
    P=\{1,a,b,ab,bs^*,\]\
    \[\forall_{i \neq n+1}\; \gamma^i,\]
    \[\forall_{i \in [1,2n], j \in [1,m^*]} \; \frac{b}{a^j}\gamma^i,\]
    \[\forall_{i \in V^*} \; s^*\gamma^{n+1-i},\]
    \[\forall_{i \in [1,2n],j \in [-m^*,m^*-1]/\{0\}} \; a^j\gamma^i\}
    \]
Also, using a new generator $g^\prime : g^\prime=g^{a^{m^*}}$ it is easy to transform each term in $P$ into monomials where the maximum degree between $f^*=b s^* \gamma^{n+1}$ and any of them is $d=2m^*+2n-1$.
Therefore, the advantage of the adversary $\epsilon=Pr[A(\beta^\prime==\beta)]$ cannot exceed $\frac{(q+2\sigma+2)^2\cdot d}{2p}$, with $\sigma$ being the (variable) number of monomials in $P$. $\square$
\end{proof}
\end{document}